\documentclass[aps,twocolumn,amsmath,showkeys,showpacs,superscriptaddress,prm]{revtex4-2}

\usepackage{dcolumn}
\usepackage{bm}
\usepackage[T1]{fontenc}
\usepackage{amsmath}
\usepackage{graphicx}
\usepackage{tabularx}

\usepackage{orcidlink}

\usepackage{hyperref}
\hypersetup{colorlinks=true, linkcolor=blue, citecolor=blue, urlcolor=blue, pdftitle={Chiral magnetism, lattice dynamics, and anomalous Hall conductivity in the novel V$_3$AuN antiferromagnetic antiperovskite}, pdfauthor={A. C. Garcia-Castro}}

\begin{document}
\title{Chiral magnetism, lattice dynamics, and anomalous Hall conductivity in the novel V$_3$AuN antiferromagnetic antiperovskite}

\author{J. M. Duran-Pinilla}
\affiliation{School of Physics, Universidad Industrial de Santander, Carrera 27 Calle 09, 680002, Bucaramanga, Colombia}

\author{Aldo H. Romero\,\orcidlink{0000-0001-5968-0571}}
\email{aldo.romero@mail.wvu.edu }
\affiliation{Department of Physics and Astronomy, West Virginia University, WV-26506-6315, Morgantown, United States}
\affiliation{Department of Physics and Materials Science, University of Luxembourg, 1511 Luxembourg, Luxembourg.}

\author{A. C. Garcia-Castro\,\orcidlink{0000-0003-3379-4495}}
\email{acgarcia@uis.edu.co}
\affiliation{School of Physics, Universidad Industrial de Santander, Carrera 27 Calle 09, 680002, Bucaramanga, Colombia}

\begin{abstract} 
Antiferromagnetic antiperovskites, where magnetically active 3$d$ metal cations are placed in the octahedral corners of a perovskite structure, are in the spotlight due to their intertwined magnetic structure and topological properties. 
Especially their anomalous Hall conductivity, which can be controlled by applied strain and/or electric field, makes them highly attractive in different electronic applications.
Here, we present the study and theoretical understanding of a new antiperovskite compound that can offer enormous opportunities in a broad set of applications.
Using first-principles calculations, we investigated the structure, lattice dynamics, noncollinear magnetic ordering, and electronic behavior in the Vanadium-based antiperovskite V$_3$AuN. 
We found an antiperovskite structure centered on N similar to the Mn$_3A$N family as the structural ground state. 
In such a phase, a \emph{Pm$\bar{3}$m} ground state was found in contrast to the \emph{Cmcm} post-antiperovskite layered structure, as in the V$_3A$N, $A$ = Ga, Ge, As, and P. 
We studied the lattice dynamics and electronic properties, demonstrating its vibrational stability in the cubic structure and a chiral antiferromagnetic noncollinear ordering as a magnetic ground state. 
Finally, we found that the anomalous Hall conductivity, associated with the topologically features induced by the magnetic symmetry, is $\sigma_{xy}$ = $-$291 S$\cdot$cm$^{-1}$ ($\sigma_{111}$ = $-$504 S$\cdot$cm$^{-1}$). The latter is the largest reported in the antiferromagnetic antiperovskite family of compounds. \\
\\
DOI:
\end{abstract}


\maketitle

\textbf{Introduction---}
The antiperovskite structure exhibits a broad and diverse set of unconventional physical and chemical properties generated by the comprehensive structural features inherited from their counterparts perovskites \cite{Wang2019, Garcia-Castro2020, Garcia-Castro2019}. 
Among them, the Mn$_3A$N family ($A$ = Ni, Cu, Zn, Ga, Ge, Pd, In, Sn, Ir, and Pt) has shown remarkable properties that cover temperature-dependent magnetic and structural transitions  \cite{Wu2013}, frustrated noncollinear ordering \cite{Fruchart1978}, large spin-phonon coupling \cite{acgarciacastro2021}, barocaloric response \cite{PhysRevX.8.041035}, strain tunability of the magnetic response \cite{Boldrin2019}, giant piezomagnetism and magnetostriction \cite{Boldrin2019,PhysRevB.78.184414, PhysRevB.96.024451, Gomonaj1989} and electrical control of their topological properties \cite{Liu2018, Tsai2020}.
All the latter features are strongly intertwined with the magnetic structure and related symmetry operations.
Regarding the magnetic response, most of the theoretical and experimentally focused efforts have been dedicated to the Mn$_3A$N family, leaving aside novel and unexplored materials with possible enhanced properties but with different magnetic cations. 
Recently, some efforts have been dedicated to the Cr$_3A$N family, showing strong magnetostructural coupling, noncollinear magnetic ordering, and tangible topologically related properties \cite{Singh2021}.
Along these lines, the vanadium-based antiperovskite family, V$_3A$N, has also been studied and some examples reported ($A$ = Ga, Ge, As, and P). 
In particular, the cases with $A$ = P and As have been experimentally analyzed and superconducting behavior was observed \cite{10.1038/srep03381}. 
Nevertheless, unlike in the Mn$_3A$N and Cr$_3A$N cases, the reported V$_3A$N materials belong to the layered orthorhombic anti-postperovskite structure \emph{Cmcm} (SG. 63)  \cite{postperovskite, https://doi.org/10.1002/pssb.201451400, doi:10.1021/ic300118d, PhysRevB.90.064113}.
This anti-postperovskite is far from the cubic symmetry shown by its Mn-based counterparts, where the crystal symmetry and magnetic ordering couple to give rise to the topological properties. 
Therefore, to expand the candidates of cubic antiperovskites, where the formation of Kagome lattices is crucial to the emergence of magnetic ordering and topological properties \cite{PhysRevLett.112.017205, Liu2018}, more efforts and studies are needed to discover novel antiperovskite candidates among other families of compounds.
For instance, it is expected that similar cubic antiperovskite compounds exist among this family in which, topological properties associated with intertwined magnetic structure and their symmetry can be present. 
Moreover, the V$_3A$N family offers an additional advantage based on its large vanadium radius ($r_V$ = 171 pm \cite{doi:10.1063/1.1712084}) at the octahedral corner site, compared to Mn ($r_{Mn}$ = 161 pm \cite{doi:10.1063/1.1712084}) potentially allowing the incorporation of larger and heavier elements as cell corner sites and maintaining the noncollinear magnetic response. The latter is highly desirable considering that strong spin-orbit coupling is a key ingredient to enlarged topologically related properties, as in the case of the anomalous Hall conductivity \cite{Singh2021}.
Furthermore, to make this type of compounds even more attractive, large topological properties and possible electrical tuning are expected for noncollinear spintronics applications \cite{Qin2019, PhysRevB.101.140405}.
As such, in this study, we theoretically propose, based on the perovskites structural criteria, a novel candidate among the antiperovskites V$_3A$N family. Herewith, we explain the atomic size relationship and tolerance conditions for an antiperovskite to hold the cubic \emph{Pm$\bar{3}$m} (SG. 221) symmetry. Furthermore, this structure can allocate a heavy metallic element as a cell corner site, providing a robust spin-orbit coupling to the Kagome lattice and enhancing the electronic properties. 
Thus, we demonstrate and explain the ground state structure by showing the structural stability from the phonon spectra, considering the antiferromagnetic noncollinear ordering.
Consequently, we explored its allowed noncollinear magnetic orderings that result in nonzero magnetic vector chirality and found that in terms of the magnetocrystalline anisotropy energy, the magnetic structure of the ground state symmetry shows a $\Gamma_{4g}$ chiral magnetic ordering. 
The latter is allowed by symmetry, and the presence of tangible Berry curvature is also demonstrated in our work. 
Finally, such a magnetic noncollinear ordering coupled to a robust spin-orbit coupling provided by the Au-sites, embedded into the Kagome lattice, induces one of the most considerable anomalous Hall conductivity obtained in the antiperovskite family of materials.

\begin{figure*}[]
 \centering
 \includegraphics[width=16.0cm,keepaspectratio=true]{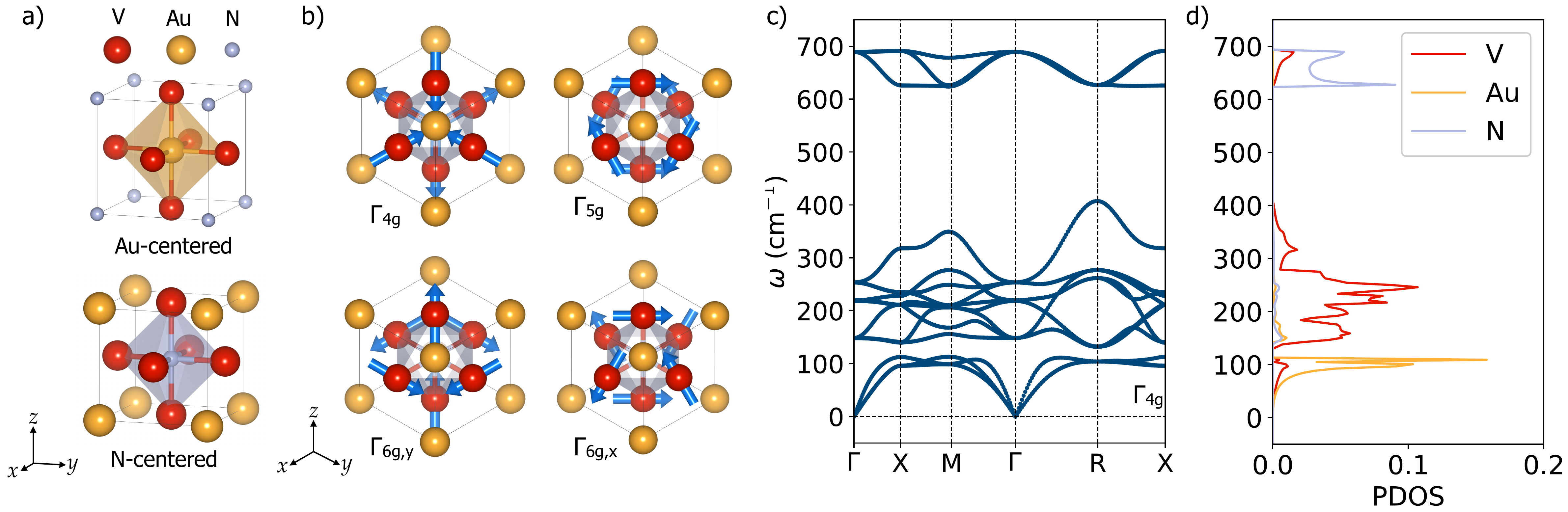}
 \caption{(Color online) (a) V$_3$AuN $Pm\bar{3}m$ structure for Au- and N-centered octahedra. (b) $\Gamma_{4g}$, $\Gamma_{5g}$, $\Gamma_{6g,y}$, and $\Gamma_{6g,x}$ noncollinear magnetic orderings allowed in the antiperovskite with the crystallographic $Pm\bar{3}m$ symmetry and labeled in agreement with  Bertaut's notation \cite{Fruchart1978,PhysRevB.101.140411}. (c) Full phonon-dispersion curves were computed at the N-centered antiperovskite considering $\Gamma_{4g}$ noncollinear ordering along the $\Gamma$$-$$X$$-$$M$$-$$\Gamma$$-$$R$$-$$X$ path in the Brillouin zone. (d) Projected phonon-DOS for the V, Au, and N sites. Here, a fully stable structure can be appreciated for this magnetic solution.}
 \label{F1}
\end{figure*} 

\textbf{Computational Details---}
We used density functional theory (DFT) \cite{PhysRev.136.B864,PhysRev.140.A1133} calculations to compute all properties reported here by using the \textsc{vasp} code (version 5.4.4) \cite{Kresse1996,Kresse1999}. The projected-augmented waves approach, PAW \cite{Blochl1994}, was used to represent the valence and core electrons. The electronic configurations considered in the pseudo-potentials as valence electrons are V: (3$p^6$3$d^4$4$s^1$, version 07Sep2000), Au: (5$d^{10}$6$s^1$, version 04Oct2007), and N: (2$s^2$2$p^3$, version 08Apr2002). 
The exchange-correlation was represented within the generalized gradient approximation (GGA-PBEsol) parametrization \cite{Perdew2008} and the V:3$d$ electrons were corrected through the DFT$+U$ approximation within the Liechtenstein formalism \cite{Liechtenstein1995}. 
We used a parameter value $U$ = 1.2 eV that was optimized to reproduce the electronic structure, and lattice parameter obtained within the meta-GGA formalism \cite{PhysRevB.84.035117} calculation in the SCAN representation \cite{PhysRevLett.115.036402}. Moreover, the magnetic moment was also compared with the obtained by means of HSE06 hybrid functional-based calculation \cite{doi:10.1063/1.2404663}. 
The periodic solution of the crystal was represented by using Bloch states with a Monkhorst-Pack \cite{PhysRevB.13.5188} \emph{k}-point mesh of 13$\times$13$\times$13 and 600 eV energy cut-off to give forces convergence of less than 0.001 eV$\cdot$\r{A}$^{-1}$ and energy less than 0.1 meV.  
Spin-orbit coupling (SOC) was included to describe the noncollinear magnetic configurations correctly \cite{Hobbs2000}. The thermal stability of the compound under different noncollinear magnetic orderings was tested by computing the phonon-dispersion curves, which were obtained using the finite-displacement approach \cite{PhysRevLett.48.406, PhysRevB.34.5065}. 
Phonon dispersions were post-processed with the \textsc{Phonopy} code \cite{phonopy}.
To obtain the anomalous Hall conductivity and Berry curvature, we used the Wannier functions methodology for which the wannierization was performed with the \textsc{Wannier90} code \cite{MOSTOFI20142309, Pizzi_2020} and post-processed with the \textsc{WannierBerri} package \cite{wannierberri}. Here, $s$, $p$, and $d$ orbitals were considered in the V and Au cases, while $s$ and $p$ were considered at the N site. Additionally, a 6.0 eV window was used around the Fermi level for the wannierization.
Finally, the atomic structure figures were elaborated with the \textsc{vesta} code \cite{vesta}.

\textbf{Results and Discussion---}
As a result of a structural search over novel antiperovskite compounds within the $M_3A$N family, we found as a potential candidate the V$_3$AuN compound \cite{V3AuN-ICSD-1965, osti_1279546}. This structural search was performed within the Materials Project database \cite{Jain2013}, and the ICSD database \cite{ICSD1,Belsky:an0615}. The performed constrain search imposed a fixed cubic \emph{Pm$\bar{3}$m} symmetry and magnetically active 3$d$ metal cations in the octahedral $M$-site added to nitrogen and 4$d$/5$d$ heavy metal cations expected to occupy the octahedral center and corner cell $A$-sites.
Fig. \ref{F1}(a) depicts the Au- and N-centered \emph{Pm$\bar{3}$m} V$_3$AuN antiperovskite structures where the octahedra are formed by the V-atoms. 
Here, in the Au-centered case, reported in Ref. \cite{osti_1279546}, the total energy for the spin-polarized FM ordering is $E$ = $-$26.907 eV$\cdot$f.u.$^{-1}$ whereas, in the former N-centered antiperovskite, the total energy in the FM ordering is $E$ = $-$40.917 eV$\cdot$f.u.$^{-1}$ representing a difference in energy of $\Delta E$ = $-$14.010 eV$\cdot$f.u.$^{-1}$ in favor of the N-centered structure.
This considerable difference in the total energy suggests the regular N-centered antiperovskite is potentially the ground state of this materials, also observed in the Mn$_3A$N family \cite{PhysRevB.100.094426, PhysRevResearch.2.023134}. 
Such a result is in apparent contradiction with the literature reported structure \cite{osti_1279546}, which claims a Au-centered type perovskite-like structure belonging to the \emph{Pm$\bar{3}$m} structure. 
However, this N-centered structure agrees with W. Rieger \emph{et al.} \cite{V3AuN-ICSD-1965} that reported the existence of V$_3$AuN belonging to the cubic perovskite symmetry with lattice parameters of $a$ = 3.962 \r{A} within the \emph{Pm$\bar{3}$m} space group \cite{V3AuN-ICSD-1965}. 
For instance, the latter structure follows the same crystalline arrangement observed in antiperovskite-like materials such as Mn$_3A$N, with $A$ = Ni, Sn, Ge, Ga, and Pt. 
Moreover, our results show that in the Au-centered antiperovskite the relaxed lattice parameter is $a$ = 5.029 \r{A}, far from the experimentally reported value of $a$ = 3.962 \r{A}.
Thus, as indicated by our results and by W. Rieger \emph{et al.} \cite{V3AuN-ICSD-1965}, such an Au-centered structure is considerably higher in energy, and the N-centered structure is potentially the crystallographic ground state.
As observed in the Mn$_3$NiN antiperovskite, it is recommended to correct the electronic exchange-correlation by including the Coulomb $U$ repulsion parameter in the magnetically active cation to adequately reproduce the vibrational, structural, and electronic degrees of freedom in the V$_3$AuN \cite{acgarciacastro2021,dtorresamaris-2022}. 
Along these lines, we have fully relaxed the electronic and atomic structure of the V$_3$AuN compound within the metaGGA SCAN-based exchange-correlation \cite{PhysRevLett.115.036402} and we obtained a relaxed lattice parameter of 3.961 \r{A} in good agreement with the experimentally reported value of $a$ = 3.962 \r{A} \cite{V3AuN-ICSD-1965}. 
An $U$ value within the PBEsol+$U$ scheme was selected to reasonably reproduce the compound's lattice parameter and electronic features. 
Based on this criterion, we selected an $U$ = 1.2 eV in the magnetically active 3$d$:V orbitals, which has been used in our calculations. Regarding the magnetic moment per vanadium atom, we obtained values of $m$= 1.134 $\mu_B$/atom and $m$= 1.256 $\mu_B$/atom for the PBEsol+$U$ and SCAN exchange-correlation representations, respectively. 
We have also compared with calculations obtained by using hybrid-functionals within the HSE06 scheme and found that, at the experimentally observed lattice parameter, the magnetic moment was $m$= 1.824 $\mu_B$/atom. Despite the latter magnetic moment being larger than the computed with the SCAN and PBEsol$+U$ schemes, all of this confirms the magnetically active behavior of the V$_3$AuN antiperovskite, and further experimentally focused efforts are needed to confirm this observable.

\begin{table}[!b]
\caption{Total energy per formula unit, in eV$\cdot$f.u.$^{-1}$, computed for the antiperovskite \emph{Pm$\bar{3}$m} and the anti-postperovskite \emph{Cmcm} structures within the V$_3A$N family, with $A$ = P, As, Sb, and Au. Here, a $U$ = 1.2 eV is considered for the V-site. Total energy values are organized following the atomic radius of the $A$-site, $r_{A}$ in pm, also presented. The energy difference is taken as $\Delta$E = E$_{Pm\bar{3}m}$ $-$  E$_{Cmcm}$, also in eV$\cdot$f.u.$^{-1}$. Goldschmidt's tolerance factor $t$, \cite{goldschmidt} was calculated considering the atomic radius as reported by E. Clementi \emph{et al.} with $r_V$ = 171 pm and $r_N$ = 56 pm \cite{doi:10.1063/1.1712084}.}
\begin{center}
\centering
\begin{tabular}{c c c c c c}
\hline
\hline
Comp. & \emph{Cmcm} & \emph{Pm$\bar{3}$m}  &   $\Delta$E & $r_{A}$  & $t$-factor \rule[-1ex]{0pt}{3.5ex} \\
\hline
V$_3$PN&  $-$44.237&  $-$42.660 &  $+$1.576 & 98 & 0.84\\ 
V$_3$AsN &  $-$42.578   &   $-$41.729 & $+$0.849 & 114  & 0.89\\
V$_3$SbN\footnote{$A$ = Sb compound is included for comparison.}&  $-$40.917 &  $-$40.816  &  $+$0.100  & 133 &  0.95 \\ 
V$_3$AuN &  $-$40.728 &  $-$40.917   &  $-$0.189 &  174 & 1.07\\
\hline
\hline
\end{tabular}
\end{center}
\label{tab:1}
\end{table}

Now, we describe the relationship between the ionic radius and the appearance of the antiperovskite and anti-postperovskites structures in the Vanadium-based family.
In the V$_3A$N group of compounds, most of the reported materials are found to crystallize in the layered anti-postperovskite \emph{Cmcm} structure that was also observed in the perovskite fluoride \cite{PhysRevB.90.064113, doi:10.1021/ic300118d, https://doi.org/10.1002/pssb.201451400} and oxide \cite{Tateno-2015,Hirose2005,Shirako2009} compounds.
We also explored the possibility for the V$_3$AuN compound to crystallize in the \emph{Cmcm} symmetry group. 
In Table \ref{tab:1} we show the calculated total energy values, per formula unit in eV$\cdot$f.u.$^{-1}$, as well as the relative energy differences in the V$_3A$N with $A$ = P, As, Sb, and Au. 
The total energies were obtained after full electronic and atomic relaxation at the cubic antiperovskite \emph{Pm$\bar{3}$m} and the anti-postperovskite \emph{Cmcm} symmetries.
As it can be observed, the energy difference is quite significant, close to $\Delta E$ = 1.576 eV, for $A$ = P. 
Thus, as soon as the corner site's atomic radius is increased, the energy difference is reduced when going from P to Sb, suggesting a possible condensation of the antiperovskite into cubic symmetry. 
Once we reach the $A$ = Au, the energy difference is $\Delta E$ = $-$0.189 eV confirming the existence of the \emph{Pm$\bar{3}$m} phase as the ground state over the layered \emph{Cmcm}.
Then, the lowest-energy structure of the compounds, except for $A$ = Au, belongs to the \emph{Cmcm} anti-postperovskite.
In perovskites, this phase is a result, in most cases, of an increase in octahedral rotations and tilts induced by isotropic pressure applied to the orthorhombic \emph{Pnma} phase.
Consequently, when the rotations overpass a limit, close to 26$^\circ$, the corner-shared octahedral lattice is broken and, to minimize the enthalpy, it converges to an edge-shared octahedra structure in which the octahedral layers are separated by $A$-site layers \cite{Okeeffe-1979}.
In other compounds, such as CaIrO$_3$ where the $A$-site radius is small when compared to the octahedra formed by IrO$_6$, the structure crystallizes at atmospheric pressure in the \emph{Cmcm} postperovskite symmetry \cite{MCDANIEL1972275, PhysRevB.76.144119}, as in the V$_3$PN and V$_3$AsN cases.
Therefore, in the anti-postperovskite structure, these results can be correlated with the corner site size, also presented in Table \ref{tab:1}, for which more significant is the $A$-site, more minor is the octahedral free space, and smaller is the tendency to observe octahedral rotations and tiltings in the antiperovskite structure. 
The latter explains the cubic \emph{Pm$\bar{3}$m} phase ground state rather than the \emph{Cmcm} layered structure in the V$_3$AuN antiperovskite.
In conclusion, the anti-postperovskite phase existence in the V$_3A$N, $A$ = P, As, and Sb is explained in terms of the tolerance factor in the perovskite structure (see Table \ref{tab:1}). 
Then, as soon as the $A$-site becomes smaller, more significant octahedral NV$_6$ rotations are expected up to the point where the crystal cannot hold the corner-shared structure, and then, a reorganization into the layered structure is expected.
Consequently, large $A$-sites in the V$_3A$N should be explored to stabilize and favor the cubic corner-shared structure.

Once we have understood the crystallographic degrees of freedom, we explore the electronic, and the allowed noncollinear magnetic structures. 
To do so, we examined the symmetry-allowed magnetic orderings, with a propagation vector \textbf{q} = (0,0,0), in the V$_3$AuN antiperovskite. 
This structure allows four different noncollinear magnetic orderings, resulting from the relaxation of magnetic frustration expected in the Kagome lattices. The latter are formed by vanadium sites in the $\langle$111$\rangle$ family of planes. 
In Fig. \ref{F1}(b) are shown the $\Gamma_{4g}$, $\Gamma_{5g}$, $\Gamma_{6g,y}$, and $\Gamma_{6g,x}$, magnetic orderings that reduce the \emph{Pm$\bar{3}$m} crystallographic symmetry group to the \emph{R$\bar{3}$m'} (MSG 166.101),  \emph{R$\bar{3}$m} (MSG 166.97), \emph{C2'/m'} (MSG 12.62), and \emph{C2/m} (MSG 12.58) magnetic symmetry groups, respectively. 
In the (111)-plane, these orderings hold a magnetic moment vector chirality of $+$1 in the $\Gamma_{4g}$ and $\Gamma_{5g}$ whereas is $-$1 in the $\Gamma_{6g,y}$, and $\Gamma_{6g,x}$ \cite{Grohol2005,PhysRevB.99.224404}.
After a complete electronic and atomic relaxation within the four noncollinear antiferromagnetic states, we obtained a total energy per formula unit of $-$41.4020 eV$\cdot$f.u.$^{-1}$, $-$41.4011 eV$\cdot$f.u.$^{-1}$, $-$41.4012 eV$\cdot$f.u.$^{-1}$, and $-$41.4015 eV$\cdot$f.u.$^{-1}$ for the $\Gamma_{4g}$, $\Gamma_{5g}$, $\Gamma_{6g,y}$, and $\Gamma_{6g,x}$ orderings, respectively. 
For comparison, we also relaxed the FM state, \emph{R$\bar{3}$m'} (MSG 166.101), with the magnetic moments pointing towards the (111)-axis. We obtained a total energy of $-$41.3066 eV$\cdot$f.u.$^{-1}$ in this magnetic state.
Thus, we estimated the energies differences with reference to the FM ordering and obtained that $\Delta E$ = $-$95.4 meV$\cdot$f.u.$^{-1}$, $-$94.5 meV$\cdot$f.u.$^{-1}$, $-$94.6 meV$\cdot$f.u.$^{-1}$, and $-$94.9  meV$\cdot$f.u.$^{-1}$ for the $\Gamma_{4g}$, $\Gamma_{5g}$, $\Gamma_{6g,y}$, and $\Gamma_{6g,x}$ orderings, respectively.
As such, in terms of the magnetocrystalline anisotropic energy, the V$_3$AuN prefers a noncollinear antiferromagnetic solution as expected from the triangular coordination in the kagome lattice and magnetic frustration of the V-sites.
The minor differences in the energy, less than 1 meV between antiferromagnetic orderings, suggest that these magnetic states barely degenerate, and they might be observed experimentally with a preference for the $\Gamma_{4g}$ order.

Aiming to explore the lattice dynamics and to confirm the vibrational stability of the V$_3$AuN compound in the cubic antiperovskite symmetry within the noncollinear magnetic orderings, we computed the phonon-dispersion curves by considering the four chiral noncollinear antiferromagnetic states.
These calculations were performed following the procedure based on the finite displacements method \cite{PhysRevLett.48.406, PhysRevB.34.5065}. Here, it is used a 2$\times$2$\times$2 supercell in which the noncollinear chiral antiferromagnetic ordering is preserved, and the displacements are considered within the \emph{R$\bar{3}$m} crystallographic symmetry. 
Later, the dynamical matrix is reconstructed and the phonon-dispersion curves are extracted. This methodology was also used for the Mn$_3$NiN antiperovskite in which a large magnetostructural coupling is observed \cite{acgarciacastro2021}.
In Fig. \ref{F1}(c), we present the full phonon-dispersion curves for the ground state $\Gamma_{4g}$, along the cubic symmetry path reference for simplicity. 
The vibrational landscape demonstrates that the N-centered cubic antiperovskite is fully stable with no negative phonons ($i.e.$ unstable modes) in the $\Gamma_{4g}$ ground state. 
Regarding the other three antiferromagnetic orderings, the phonon dispersion curves also show fully dynamically stable structures, as shown in Fig. \textcolor{blue}{S1} with no major difference for those antiferromagnetic orderings. 
In Fig. \ref{F1}(d) is presented the projected phonon-DOS extracted for the $\Gamma_{4g}$ order. As it can be observed, the phonons around 100 cm$^{-1}$ are strongly dominated by Au sites with a slight overlap with the V sites because of the small contribution to Au-driven eigen-displacements. 
For frequencies between 140 cm$^{-1}$ to 400 cm$^{-1}$, we observe the vibrational modes dominated by the V-sites eigen-displacements. 
Such atomic vibrations are associated with octahedral rotations and tiltings related to the Kagome lattice. 
Finally, the N-site's phonons are located between 600 cm$^{-1}$ to 700 cm$^{-1}$ well above at large frequency values. 
We found a band gap in the phonon dispersion between acoustic and optical modes, located close to 120 cm$^{-1}$. 
The latter phononic band gap prevents the propagation of mechanical waves at a defined frequency range. This is importantly observed in phononic crystals \cite{10.1115/1.4026911,phononic-crystal2017}, 2D \cite{doi:10.1063/1.4896685}, and 3D monochalcogenides \cite{PhysRevB.100.054104}.
As such, the phononic band gap suggests the prohibited propagation frequencies between 113 cm$^{-1}$ and 140 cm$^{-1}$, taken at the $X$-point in the BZ and potentially affecting the thermal properties, such as thermal conductivity, among others.
This band gap can be potentially attributed to the significantly different eigen-displacements below the band gap, driven by antipolar displacements of the Au-site along the $z$-axis and above, dominated by the motion of the NV$_6$ octahedral in the $xy$-plane. 
The latter is enhanced by the large mass difference between the Au sites, $w_{Au}$ = 196.9665 amu, and the V and N sites with $w_V$ = 50.9415 amu and $w_N$ = 14.0067 amu, respectively, as suggested by U. Argaman \emph{et al.} \cite{PhysRevB.100.054104}.
Nonetheless, the specific details of the physical source of the phononic band gap, and its potential influence on the thermodynamical observables, are to be studied in further work.

\begin{figure}[b]
 \centering
 \includegraphics[width=8.7cm,keepaspectratio=true]{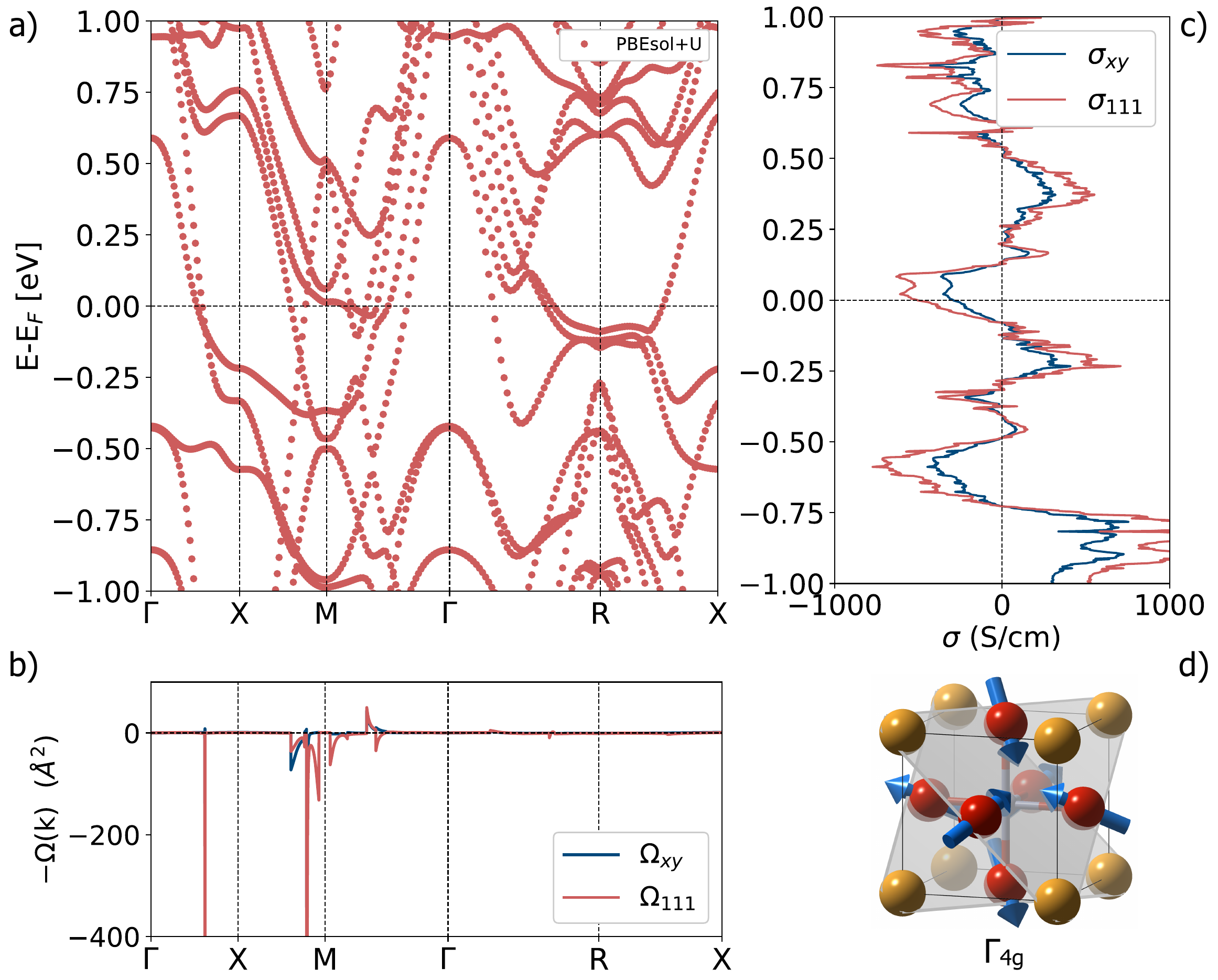}
 \caption{(Color online) (a) Electronic bands computed for the $\Gamma_{4g}$ ordering in the V$_3$AuN along the $\Gamma$$-$$X$$-$$M$$-$$\Gamma$$-$$R$$-$$X$ path in the Brillouin zone. In (b) it is presented the computed Berry curvature, $\Omega_{xy}$($\textbf{k}$) and $\Omega_{111}$($\textbf{k}$), along the same $k$-path selected for the electronic bands. (c) Anomalous Hall conductivity, $\sigma_{xy}$ and $\sigma_{111}$ conductivity components, were computed for the $\Gamma_{4g}$ ordering (d) $\Gamma_{4g}$ magnetic state in which, the magnetic moments at the Vanadium atoms are denoted by blue arrows.}
 \label{F2}
\end{figure} 

Moving forward, in Fig. \ref{F2}(a), we present the computed electronic band structure across the high-symmetry points of the cubic reference for the $\Gamma_{4g}$ state. 
Here, we observe the expected metallic behavior with a predominant contribution from the V-sites close to the Fermi energy.
Afterward, and considering the magnetic symmetry groups, we observe that the ground state antiferromagnetic $\Gamma_{4g}$ ordering allows for the appearance of non-zero Berry curvature and, therefore, net components of the anomalous Hall conductivity, AHC, tensor \footnote{The anomalous Hall conductivity component, $\sigma_{xy}$ for example, has been computed by the formula:
 \begin{align}\label{eq:ahc}
   \sigma_{xy}=-\frac{e^2}{\hbar}\int_{BZ}\frac{d^3\bf{k}}{(2\pi)^3}\Omega_{xy}(\bf{k}),
\end{align}
where $\Omega_{xy}(\bf{k})$=$\sum_n f_n(\bf{k})$$\Omega_{n,xy}(\bf{k})$ is the summation of all the included $n$ number of bands and $f_n\bf(k)$ is the Fermi distribution.
Moreover, the symmetry-allowed AHC components, within the $\Gamma_{4g}$ magnetic ordering, are:
\begin{align}\label{ahc:tensor}
\sigma_{\Gamma_{4g}}=
\begin{pmatrix}
0 & \sigma_{xy} & -\sigma_{xy}\\
-\sigma_{xy} & 0 & \sigma_{xy} \\
\sigma_{xy} & -\sigma_{xy}& 0
\end{pmatrix}
\end{align}}.
The latter is also allowed in the $\Gamma_{6g,x}$ and $\Gamma_{6g,y}$ magnetic orderings with different tensors
\footnote{The AHC tensors in the $\Gamma_{6g,y}$ and $\Gamma_{6g,x}$ symmetries are:
\begin{align}\label{ahc:tensor}
\sigma_{\Gamma_{6g,y}}=
\begin{pmatrix}
0 & \sigma_{xy} & -\sigma_{xy}\\
-\sigma_{xy} & 0 & \sigma_{xy} \\
\sigma_{xy} & -\sigma_{xy}& 0
\end{pmatrix}\\
\sigma_{\Gamma_{6g,x}}=
\begin{pmatrix}
0 & 0 & -\sigma_{zx}\\
0 & 0 & -\sigma_{zx} \\
\sigma_{zx} & \sigma_{zx}& 0
\end{pmatrix}
\end{align}}.
As such, in Fig. \ref{F2}(b) and in Fig. \ref{F2}(c) are presented the Berry curvature, $\Omega_{xy}$($\textbf{k}$) and $\Omega_{111}$($\textbf{k}$), and the anomalous Hall conductivity, $\sigma_{xy}$ and $\sigma_{111}$ components, respectively. These were computed at the $\Gamma_{4g}$ magnetic ordering ground state, also included in Fig. \ref{F2}(d).
The Berry curvature and AHC components within the [111]-plane were computed such as $\Omega_{111}($\textbf{k}$)  \equiv \frac{1}{\sqrt{3}}\left[\Omega_{xy}(\textbf{k}) +\Omega_{yz}(\textbf{k})+\Omega_{zx}(\textbf{k})\right]$ and  $\sigma_{111} \equiv \frac{1}{\sqrt{3}}\left(\sigma_{xy}+\sigma_{yz}+\sigma_{zx}\right)$, respectively.
We observed that the $\Omega_{111}$($\textbf{k}$) component, as well as the $\Omega_{xy}$($\textbf{k}$), have considerable net values along the $\Gamma$$-$$X$, $X$$-$$M$, and $\Gamma$$-$$M$ paths with divergent behavior in the $\Gamma$$-$$X$, $X$$-$$M$. The latter is due to the presence of bands crossing associated with Weyl points close to the Fermi energy, as also observed in other antiperovskites \cite{PhysRevB.100.094426}.
After calculating the $\sigma_{xy}$  ($\sigma_{111}$) component at the Fermi energy, we obtained a value of $\sigma_{xy}$ = $-$291 S$\cdot$cm$^{-1}$ ($\sigma_{111}$ = $-$504 S$\cdot$cm$^{-1}$) 
This AHC value obtained in the V$_3$AuN is, to date, among the largest observed in the antiferromagnetic antiperovskites family. 
This considerable value is explained in terms of the large spin-orbit coupling of the Au sites transferred to the Kagome lattice through hybridization with V-sites \cite{PhysRevLett.112.017205}.
Here, it is important to mention that according to Huyen \emph{et al.} \cite{PhysRevB.100.094426}, the largest theoretically reported AHC value has been calculated in the Mn$_3$PtN antiperovskite with $\sigma_{xy}$ = 462 S$\cdot$cm$^{-1}$ ($\sigma_{111}$ = 800 S$\cdot$cm$^{-1}$) \cite{PhysRevB.100.094426}, nevertheless, in our calculations we obtained an $E_{MAE}$ = $E_{\Gamma_{4g}}$ $-$ $E_{\Gamma_{5g}}$ = 1.9 meV for the Mn$_3$PtN compound. This suggests the $\Gamma_{5g}$ as the magnetic ground state in which the AHC is forbidden by symmetry. 
This finding also agrees with the results of H. K. Singh \emph{et al.} \cite{Singh2021}. 
Furthermore, at this point, it is worth commenting that in most of the reported literature, the calculations have been performed within the PBE approach, and no Hubbard term has been considered; and therefore, a no electronic correction has been explored, aiming to remove the computational chemical pressure and to obtain fully relaxed lattice parameters \cite{PhysRevResearch.2.023134,PhysRevB.100.094426,PhysRevMaterials.3.094409,PhysRevMaterials.3.044409}. 
Thus, the exchange-correlation energy, which strongly influences the magnetic and electronic structure, has not been included considering the Coulomb repulsion value $U$, which, based on our analysis, tends to considerably reduce the bands spreading and, therefore, affects the electronic structure close to the Fermi level and reduces the computed AHC values \cite{dtorresamaris-2022, acgarciacastro2021}. 
Moreover, this exchange-correlation correction affects the structural parameters where the experimentally observed lattice parameter can be obtained after carefully introducing the PBEsol$+U$ methodology, as discussed above.
After including the value of $U$ = 2.0 eV in the 3$d$:Mn orbitals for the Mn$_3$PtN compound, fitted to reproduce the experimentally obtained lattice parameter within the PBEsol+$U$ scheme, we found that $\sigma_{xy}$ = 144 S$\cdot$cm$^{-1}$ ($\sigma_{111}$ = 249 S$\cdot$cm$^{-1}$) in the $\Gamma_{4g}$ higher in energy and metastable magnetic ordering.
Evidently, this value is considerably smaller than the reported one.
For example, in the Mn$_3$NiN case, D. Torres-Amaris \emph{et al.} \cite{dtorresamaris-2022} showed that under the PBEsol$+U$ scheme, as shown here, the AHC is $\sigma_{xy}$ = 68 S$\cdot$cm$^{-1}$, in better agreement with the experimentally reported value of $\sigma_{xy}$ =  22 S$\cdot$cm$^{-1}$ in thin-films \cite{PhysRevMaterials.3.094409}, and considerably smaller than the calculated value of $\sigma_{xy}$ = 130--170 S$\cdot$cm$^{-1}$ in Refs. \cite{PhysRevMaterials.3.044409, PhysRevMaterials.3.094409} where no correction was considered.
Moreover, the PBEsol failure to reproduce the experimental lattice parameter is solved by including $U$ to correctly produce the spin-phonon coupling in these antiperovskites. 
The validation with the metaGGA SCAN calculations indicates that the exchange-correlation needs to be treated carefully in this type of material. 
As such, we have also computed the AHC component within the SCAN formalism in the V$_3$AuN antiperovskite and we found that $\sigma_{xy}$ = $-$347 S$\cdot$cm$^{-1}$ ($\sigma_{111}$ = $-$600 S$\cdot$cm$^{-1}$), in fair agreement with the values obtained by the PBEsol+$U$ approach.
Furthermore, it is worth commenting that SCAN-based calculations have also shown a considerable improvement in describing the electronic and structural behavior in Mn-rich compounds \cite{PhysRevB.101.075115}, Heusler alloys \cite{PhysRevB.102.045127}, metal oxides \cite{PhysRevMaterials.2.095401}, and in the Mn$_3$NiN antiperovskite \cite{acgarciacastro2021}.
Even though several reports suggest an overestimation of the magnetic moment when the exchange-correlation SCAN is used. In the case of antiperovskites, the total magnetic moment is zero due to the antiferromagnetic ordering ground state. Moreover, there is no suggested connection between the magnetic moment per atom and the AHC, which here only depends on the symmetry operations associated with the overall antiferromagnetic noncollinear ordering.
Therefore, a future detailed work dedicated to exploring the dependence of the electronic and magnetic structure on the exchange-correlation approach in these and other magnetically-driven topological materials is highly desirable. 
Notably, although the antiferromagnetic chiral structure is responsible for the symmetry conditions that allow the appearance of net Berry curvature and topological features, the allocation of 5$d$ heavy atoms in the lattice, such as Au in this case, considerably enhances the AHC values due to their sizeable spin-orbit coupling when compared to 4$d$ and 3$d$ metals \cite{PhysRevLett.112.017205,PhysRevB.100.094426,Singh2021}.
Finally, we believe that our work opens the possibility for the existence of other antiperovskites compounds, such as V$_3$AgN and V$_3$BiN, where the Bi's and Ag's large atomic size could induce an N-centered cubic antiperovskite structure. In addition to the expected sizeable spin-orbit coupling interaction that can give rise to large AHC and unexpected properties associated with their topological features.
Our results show that closed shell elements, such as $A$ = Ag and Bi, retain the magnetism in the V$_3A$N compounds, whereas open shell cations, such as $A$ = Pt and Ir, induce a nonmagnetic state. This observation calls attention to further studies aiming to investigate in more detail the effect of the $A$-site on the electronic properties among all the antiperovskite compounds.

\textbf{Conclusions---}
We have investigated, for the first time and using first-principles calculations, the novel Vanadium-based antiperovskite nitride V$_3$AuN. 
Despite the reported Au-centered octahedral antiperovskite structure, the N-centered structure is potentially the ground state energy structure. It is supported by the fact that it is also dynamically stable, as demonstrated by the computed phonon-dispersion curves. 
Additionally, the V$_3$AuN is the first reported vanadium-based antiperovkite that belongs to cubic symmetry rather than the layered post-antiperovskite materials reported among the V$_3A$N family.
The cubic structure entirely agrees with other Mn- and Cr-based antiperovskites compounds.
As observed, four different noncollinear magnetic orderings are allowed resulting from the magnetic frustration in the kagome lattices formed by the vanadium sites in the $\langle$111$\rangle$ family of planes.
Here, we observed that the chiral $\Gamma_{4g}$ is the lowest energy magnetic ground state in which a non-zero Berry curvature and, consequently, a tangible anomalous Hall conductivity is present.
Interestingly, when the exchange-correlation correction is considered, and the correct magnetic ground state is taken into account in the Mn$_3$PtN compound, the computed $\sigma_{xy}$ and $\sigma_{111}$ values in the V$_3$AuN are among the largest reported in the antiperovskite family of materials, of course, waiting for experimental confirmation. 
Finally, we hope our results will motivate more studies in these novel antiferromagnetic antiperovskite families of materials. Those with several topological signatures that couple to the fascinating magnetic response can be explored and possibly controlled for future antiperovskite spin-based devices.


\textbf{Acknowledgements---}
Calculations presented in this article were carried out using the GridUIS-2 experimental testbed, being developed under the Universidad Industrial de Santander (SC3-UIS) High Performance and Scientific Computing Centre, development action with support from UIS Vicerrector\'ia de Investigaci\'on y Extensi\'on (VIE-UIS) and several UIS research groups as well as other funding resources.
We also acknowledge the computational resources awarded by XSEDE, a project supported by National Science Foundation grant number ACI-1053575. The authors also acknowledge the support from the Texas Advances Computer Center (with the Stampede2 and Bridges supercomputers). Additionally, the Super Computing System (Thorny Flat) at WVU, which is funded in part by the  National Science Foundation (NSF) Major Research Instrumentation Program (MRI) Award \#1726534, and West Virginia University is also recognized. A.C.G.C. acknowledge the grant No. 2677 entitled “Quiralidad y Ordenamiento Magnético en Sistemas Cristalinos: Estudio Teórico desde Primeros Principios” supported by the VIE – UIS.

\bibliography{library}

\end{document}